\documentclass[fleqn,usenatbib]{mnras}

\usepackage{footnote}
\usepackage[T1]{fontenc}
\usepackage{ae,aecompl}
\usepackage{longtable}
\usepackage{longtable}
\usepackage{graphicx}
\usepackage{wrapfig}
\usepackage{lscape}
\usepackage{rotating}
\usepackage{epstopdf}

\usepackage{graphicx}	
\usepackage{amsmath}	
\usepackage{amssymb}	
\newcommand{\kms}{km\,s$^{-1}$} 
\newcommand{\HI}{H{\sc i }}

\newcommand{\hi}{H{\sc i}}
\newcommand{\hii}{H{\sc i}~21\,cm}

\title[H{\sc i} 21\,cm absorption in GPS sources]{A Giant Metrewave Radio Telescope search for associated H{\sc i} 21\,cm 
absorption in GHz-peaked-spectrum sources.}

\author[Aditya \& Kanekar]{J. N. H. S. Aditya$^{1,2}$\thanks{adityaj@iucaa.in},
Nissim~Kanekar$^2$\thanks{Swarnajayanti Fellow; nkanekar@ncra.tifr.res.in}\\
$^{1}$Inter-University Centre for Astronomy and Astrophysics, Pune 411007, India\\
\newline
$^{2}$National Centre for Radio Astrophysics, Tata Institute of Fundamental Research, Pune 411007, India}

\date{Accepted XXX. Received YYY; in original form ZZZ}

\pubyear{2016}

\begin{document}
\label{firstpage}
\pagerange{\pageref{firstpage}--\pageref{lastpage}}
\maketitle

\begin{abstract}
We report the first detections of associated H{\sc i} 21\,cm absorption in Gigahertz-peaked-spectrum (GPS) sources
at high redshifts, $z > 1$, using the Giant Metrewave Radio Telescope (GMRT). Our GMRT search for associated 
H{\sc i} 21\,cm absorption in a sample of 12 GPS sources yielded two new detections of absorption, towards 
TXS~1200+045 at $z = 1.226$ and TXS~1245$-$197 at $z = 1.275$, and five non-detections. These are only the 
sixth and seventh detections of associated H{\sc i} 21\,cm absorption in active galactic nuclei (AGNs) 
at $z > 1$. Both H{\sc i} 21\,cm absorption profiles are wide, with velocity spans between nulls of 
$\approx 600$~km~s$^{-1}$ (TXS~1200+045) and $\approx 1100$~km~s$^{-1}$ (TXS~1245$-$197). In both 
absorbers, the large velocity spread of the absorption and its blueshift from the AGN, suggests that 
it arises in outflowing neutral gas, perhaps driven by the radio jets to high velocities. We derive 
mass outflow rates of ${\dot M} \approx 32 \; {\rm M}_\odot$~yr$^{-1}$ (TXS~1200+045) and 
${\dot M} \approx 18 \; {\rm M}_\odot$~yr$^{-1}$ (TXS~1245$-$197), comparable to the mass outflow 
rates seen earlier in low-redshift active galactic nuclei.  

\end{abstract}



\section{Introduction}

Over the last decade, it has become clear that active galactic nuclei (AGNs) 
play a critical role in the evolution of their host galaxies and environments 
\citep[e.g.][]{fabian2012}. Simulations have shown that interactions between 
the AGN jets and the surrounding gas can regulate the growth of the host galaxy, 
by quenching star formation in the central regions and possibly ending the 
active state of the nucleus, through the effects of mechanical feedback 
\citep[e.g.][]{hopkins2005,springel2005,croton2006}. Recent simulations have shown
that the kinetic interaction of the radio jet with interstellar gas in the AGN host 
galaxy provides an efficient mechanism to drive gas outflows, especially when the 
radio source is in its initial phase and surrounded by a porous clumpy medium 
\citep[e.g.][]{wagner2011,wagner2012}. This is because a large cocoon of disturbed 
and outflowing gas is created around the radio jet due to its interaction with 
the radio plasma, thus affecting a large region of the galaxy. AGN-driven galactic winds 
provide an alternate efficient mechanism of driving gas flows away from the central 
region and regulating star formation in the host galaxy \citep[e.g.][]{veilleux05}. 
However, despite its acknowledged importance, the ubiquity of AGN feedback, its effects 
on galaxy evolution, whether the feedback arises during specific phases in the AGN lifetime 
or is a recurrent phenomenon, and the driving mechanism of the outflows all remain 
open issues today \citep[][]{fabian2012}. 

The above feedback mechanisms involve gas flowing outwards from the AGN. Conversely, 
gas flows towards the central regions of the AGN host galaxy are an important source of 
fuel for the AGN activity; the absence of such fuel can result in the cessation of 
AGN activity. The kinematics of gas in AGN environments thus has important implications 
for galaxy evolution. 

``Associated'' \hii\ absorption studies provide an interesting probe of the 
presence and kinematical properties of neutral hydrogen (\HI) in AGN environments 
\citep[e.g.][]{morganti2012}. The detection rate of \hii\ absorption at different 
redshifts provides information about the availability of fuel for the AGN activity. 
In the case of detections of \hii\ absorption, the line width, the line shape, and 
the velocity offset relative to the AGN redshift can be used to probe conditions 
in the AGN environment \citep[e.g.][]{gereb2015}. Absorption profiles that are narrow, 
with widths $\lesssim 200$~km~s$^{-1}$, are likely to arise from clouds that are 
rotating in disks around the nucleus, while broader line profiles, with widths 
$\gtrsim 300$~km~s$^{-1}$, suggest the presence of highly unsettled gas, that may
be interacting with the nuclear jets \citep[e.g.][]{gereb2015}. An absorption line 
that is redshifted from the AGN's systemic velocity is likely to indicate neutral 
gas that is flowing towards the central regions, i.e. a source of fuel for the central 
AGN \citep[e.g.][]{vangorkom89}. Conversely, absorption lines that are blueshifted 
from the systemic velocity are likely to indicate outflowing gas, due to either 
stellar outbursts or ram pressure from the AGN jets. 

The detection of blueshifted and wide associated \hii\ absorption features can be used 
to probe jet-cloud interactions, and thus aid in understanding AGN-driven feedback. 
It is also interesting to test whether there is a preponderance of blueshifted or 
redshifted absorption features at a given redshift, and whether this evolves with redshift.
Early studies of associated \hii\ absorption in the local Universe found the absorption
to lie either at or redward or the AGN systemic velocity, indicating infall of 
neutral gas to the inner regions of the AGN host galaxy, with infall rates sufficient 
to fuel the nuclear activity \citep{vangorkom89}. More recently, a number of blueshifted 
associated \hii\ absorption lines have been detected at low and intermediate redshifts, 
$z < 1$, suggesting that AGN outflows play an important role in the kinematics here 
\citep[e.g.][]{vermeulen2003,gereb2015}. Indeed, very fast \hi\ outflows, with 
velocities $\gtrsim 1000$~km~s$^{-1}$, have been observed in a few low-redshift 
AGNs \citep[IC~5063, 4C12.50, 3C293, Mrk231, etc.; e.g. ][]{morganti98,morganti05,morganti16}. 
Very Long Baseline Interferometry (VLBI) mapping in the \hii\ absorption line has been used to find 
evidence that the radio jet is driving the \hi\ outflow in a few cases \citep[IC~5063, 
4C12.50, 3C305, Mrk231; e.g. ][]{oosterloo00,morganti03}. Equally interesting, 
some of these outflows \citep[e.g. NGC~1266, Mrk231, NGC~1433, etc.; e.g. ][]{alatalo11,combes13,dasyra16} 
have been shown to have a molecular counterpart; this has important implications for their impact on both 
quenching of star formation in the host galaxy and the inter-galactic medium, as the 
presence of molecular gas implies that the outflows carry even more mass than in the 
purely atomic case \citep[e.g.][]{fiore17}.

Finally, while a number of attempts have been made to use assocatied \hii\ absorption 
studies to probe AGN environments at high redshifts, $z \gg 1$ 
\citep[e.g.][]{gupta2006,curran2013,aditya2016}, these have mostly been unsuccessful. 
There are at present only five known associated \hii\ absorbers at $z > 1$, at 
$z \approx 1.2$ towards 3C190 \citep[][]{ishwar2003} and TXS~1954+513 
\citep[][]{aditya2017}, $z \approx 1.3$ towards J1545+4751 \citep[][]{curran2013}, 
$z \approx 2.6$ towards MG~J0414+0534 \citep[][]{moore1999}, and $z \approx 3.4$ 
towards TXS~0902+343 \citep[][]{uson1991}. Indeed, the detection fraction of associated \hii\ 
absorption in compact flat-spectrum sources has been shown to have a strong dependence 
on both redshift and AGN ultraviolet/radio luminosity, with low detection rates at high
redshifts and high luminosities \citep[][]{aditya2016}. It has not so far been 
possible to separate between redshift evolution and AGN luminosity as the primary 
cause of the above effect.

At low redshifts, the highest detection rate of associated \hii\ absorption has been
found in Gigahertz Peaked Spectrum (GPS) sources \citep[e.g.][]{gupta2006}. GPS sources 
are known to be extremely compact, with transverse sizes $\lesssim 1$~kpc \citep[][]{odea98}. 
It is now widely agreed that these sources correspond to the early stages of the 
evolution of powerful radio galaxies \citep[e.g.][]{fanti1995,readhead1996,snellen2000}. 
The radio-emitting region grows and expands within the interstellar medium of the host
galaxy, before breaking out to become a powerful radio source. GPS sources thus appear to 
be good candidates to search for jet-cloud interactions, since such effects are believed 
to be prominent in young AGNs, where the radio emission is engulfed in the ambient gas 
reservoir. Further, the high detection rate of \hii\ absorption suggests that these 
may be the best targets to extend studies of associated \hii\ absorption to high redshifts,
$z > 1$.

At present, only three of the 23 GPS sources that have been searched for associated 
\hii\ absorption lie at $z > 1$. \citep[e.g.][]{vermeulen2003,gupta2006}, with no reported 
detections. The kinematical properties of neutral gas in the vicinity of high-$z$ 
GPS sources, and the redshift evolution of their environment, thus remain open 
questions. We have hence put together a sample of 58 GPS sources at declinations 
observable with the Giant Metrewave Radio Telescope (GMRT), and whose \hii\ line 
frequencies redshift into the GMRT bands, from the GPS literature 
\citep[e.g.][]{stanghellini1998,labiano2007,vries2007,randall2011}. The selection 
criteria are that the sources should have inverted spectra, with the turnover frequency 
lying between 300~MHz and 5~GHz \citep[the definition of a GPS source; e.g.][]{labiano2007}. 
23 of the 58 sources have earlier searches for \hii\ absorption in the literature. 
In this paper, we report results from a GMRT search for redshifted \hii\ absorption in 
the 12 brighest GPS sources of this sample, that has yielded the first two detections of 
\hii\ absorption in GPS sources at $z > 1$.
 
\begin{figure*}
\includegraphics[scale=0.4]{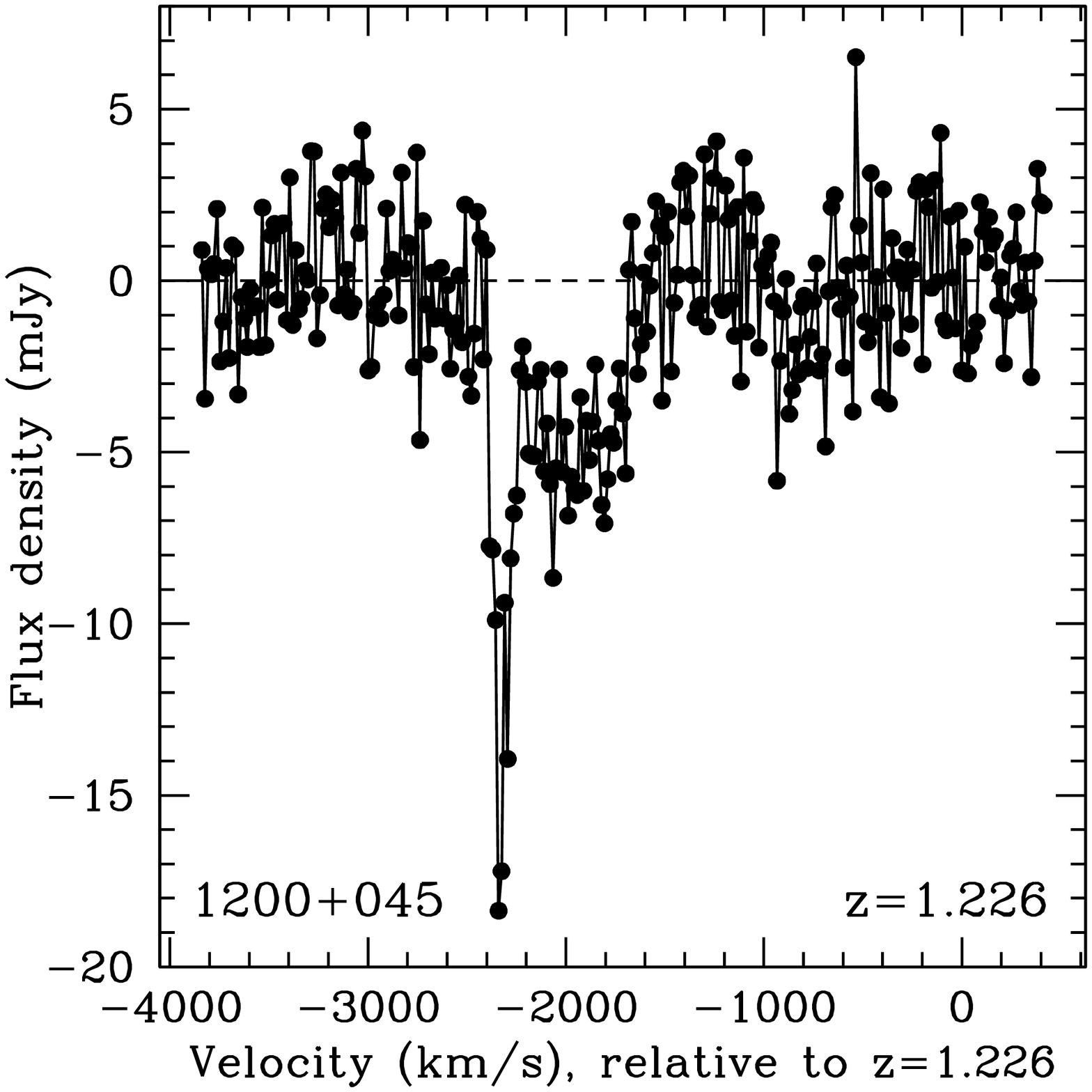}
\includegraphics[scale=0.4]{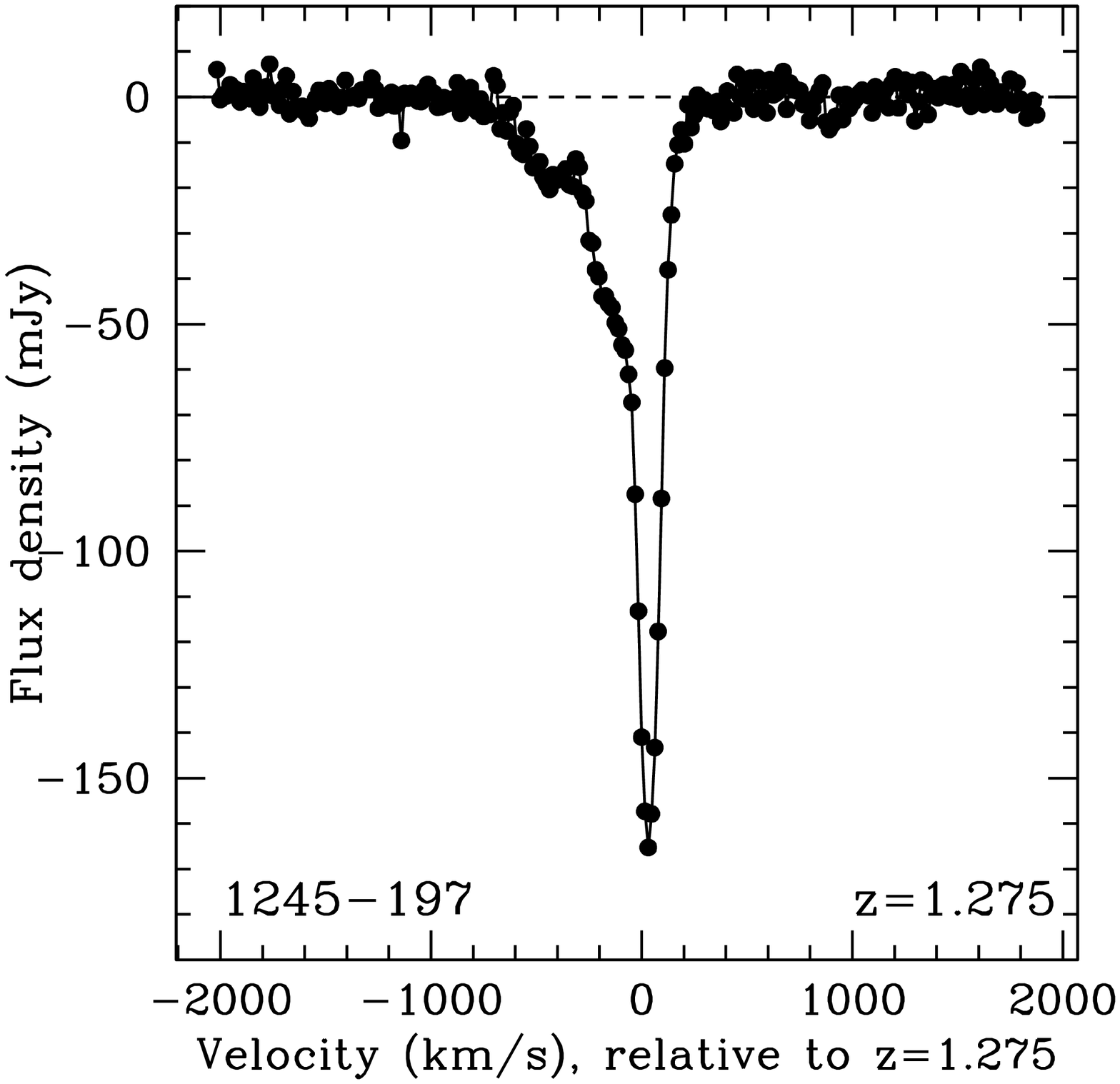}
\caption{[A]~(left panel): The GMRT~\hii\ absorption spectrum towards TXS~1200+045, 
at $z = 1.226$, from August 2016. [B]~(right panel: The GMRT~\hii\ absorption spectrum towards 
TXS~1245$-$197, at $z = 1.275$, from August 2016 and June 2017.}
\label{fig:detect}
\end{figure*}

\section{Observations, Data Analysis and Results}

\subsection{The GMRT observations}
\label{gmrt_obs}

Our GMRT search for associated \hii\ absorption from the 12 brightest GPS sources of our 
full sample was carried out in GMRT Cycle~29 over 2015 October -- December. The GMRT L-band 
receivers were used for the nine targets at $z \lesssim 0.4$, and the 610~MHz receivers for 
the three targets at $z \approx 1.1-1.5$. The GMRT Software Backend (GSB) was used as the 
correlator for all observations, with a bandwidth of 16.7~MHz centred at the expected 
redshifted \hii\ line frequency and sub-divided into 512 channels. This yielded a 
velocity coverage of $\approx 3900-5000$~\kms\ and a velocity resolution of $\approx 7.7-9.8$~\kms,
for the sources at $z < 0.4$, and a velocity coverage of $\approx 7600 - 8000$~\kms\ 
and a velocity resolution of $\approx 14.8-15.6$~\kms\ for the sources at $z \approx 1.1-1.5$.
Observations of the standard primary calibrators 3C48, 3C147 and/or 3C286 were 
used to calibrate the flux density scale and the system passband, while observations
of nearby compact sources were used for phase (and, in some cases, passband) calibration. 
The on-source times were $\approx 1$~hour apiece. 

The initial spectra of two sources, TXS~1200+045 and TXS~1245$-$197, showed tentative detections of 
\hii\ absorption. These sources were hence re-observed in 2016 August, with a similar observing 
setup (i.e. the GSB as the correlator, a bandwidth of 16.7~MHz and 512~channels, and standard flux 
density, passband and phase calibration), but with on-source times of $\approx 2$~hours apiece. 
In both cases, however, the observing band was centred at the line frequency of the 
putative absorption feature, i.e. $\approx 644.33$~MHz for TXS~1200+045 and $\approx 624.35$~MHz
for TXS~1245$-$197.

Finally, the new GMRT spectrum of TXS~1245$-$197 showed evidence of wide \hii\ absorption, in addition 
to the strong narrow component seen in the original spectrum. We hence re-observed this source in 
2017 June, to confirm the wide feature, with the same observing setup and procedure as that in 2016 
August, but with a total on-source time of 4 hours.

\begin{table*}
\begin{center}
\footnotesize
\caption{The 12 GPS sources of this paper, in order of increasing redshift.}
\label{tab:gps_12}
\begin{tabular}{ccccccccc}
\hline
Name &$z$ & $\nu_{21~cm}$ $^{a}$ & Vel. & Res. & $S_{\nu}$ & $\Delta S$ & $\int \tau d$V $^{b}$  & $N_{\rm HI}$ $^{b}$\\ 
     &    &                      & cov. &  & & & & $\times 10^{21}$ \\ 
     &    &  MHz               & km/s &  km/s & mJy & mJy & km $s^{-1}$ &  ${\rm cm}^{-2}$ \\
\hline
B3 0801+437  & 0.123 & 1264.8 & 3942.4 & 15.4 & $407.7 \pm  0.1$ & 1.42 & $< 0.58$ & $< 1.1$\\ 
TXS 1540-077 & 0.172 & 1211.9 & 4147.2 & 16.2 & 1525.9$^d$ &-- &-- &-- \\ 
TXS 0320+053 & 0.179 & 1204.7 & 4147.2 & 16.2 & 3159.9$^d$ &-- &-- &-- \\ 
TXS 1819+671 & 0.221 & 1163.3 & 4300.8 & 16.8 & $434.1 \pm  0.2$ & 1.74 & $< 0.85$ & $< 1.5$\\ 
TXS 1151-348 & 0.258 & 1129.0 & 4454.4 & 17.4 & 4650.6$^d$ &-- &-- &-- \\ 
TXS 1108+201 & 0.299 & 1093.4 & 4556.8 & 17.8 & $ 1448.7 \pm 0.2 $ & 3.93 & $< 0.44$ & $< 0.81$  \\ 
TXS 0019-000 & 0.305 & 1088.4 & 4608.0 & 18.0 & 2985.7$^d$ &-- &-- &--\\ 
TXS 0240-217 & 0.314 & 1080.9 & 4659.2 & 18.2 & 1229.9$^d$ &-- &-- &--\\ 
TXS 0507+179 & 0.416 & 1003.1 & 5017.6 & 19.6 & $504.2 \pm  0.3$ & 1.11 & $< 0.41$ & $< 0.75$\\ 
TXS 2121-014 & 1.158 & 658.2  & 7628.8 & 29.8 & $2095.2 \pm  0.4$ & 3.56 & $< 0.38$ & $< 0.69$\\ 
TXS 1200+045 & 1.226 & 638.1  & 7833.6 & 15.3$^{c}$  & $1675.2 \pm  0.4$ & 1.98 & $2.52 \pm 0.12$ & $4.59 \pm 0.22$\\ 
TXS 1245$-$197 & 1.275 & 624.3  & 7987.2 & 15.6$^{c}$ & $8302.2 \pm 0.6$ & 3.62 & $4.542 \pm 0.043$ & $8.280 \pm 0.078$   \\
\hline
\end{tabular}
\end{center}
\vskip 0.05in
Notes to the table:\\
$^a$~$\nu_{21~cm}$ is the redshifted \hii\ line frequency.\\
$^{b}$~For \hii\ non-detections, the $3\sigma$ upper limits on the velocity-integrated 
\hii\ optical depth and the \hi\ column density assume a line FWHM of 100 km~s$^{-1}$ and a 
spin temperature of 1000~K. The $3\sigma$ optical depth limits have been computed after smoothing
each spectrum to a resolution of 100~km~s$^{-1}$.\\
$^{c}$ The quoted velocity resolution is without Hanning-smoothing and re-sampling.\\
$^{d}$ The flux density of the source at the redshifted \hii\ line frequency was estimated by 
interpolating between the 1.4~GHz (from the FIRST or NVSS surveys; \citealp[][]{becker1995}; \citealp[][]{condon1998}) 
and the 325/365~MHz (from the WENSS or Texas surveys; \citealp[][]{douglas1996}; \citealp[][]{rengelink1997}) 
flux densities from the literature.
\end{table*}

\subsection{Data analysis}\label{analysis}
The GMRT data were analysed in ``classic'' AIPS, using standard procedures. The data were first 
inspected and edited, to remove non-working antennas and time-specific bad data, usually arising due 
to intermittent radio frequency interference (RFI). The antenna bandpasses were determining using 
the data on the flux calibrators. After calibration of antenna-dependent gains and bandpass shapes,
an iterative self-calibration procedure was followed for each target source. Typically, this 
consisted of 3--4 rounds of phase-only self-calibration and imaging, followed by 1--2 rounds of 
amplitude-and-phase self-calibration and imaging. At the end of the above procedure, the visibility 
data were again inspected, and further edited to remove any corrupted data, after which the self-calibration
was repeated. This iterative procedure was carried out until it yielded an image that did not improve 
upon further self-calibration and data editing. The final continuum image was then subtracted out from 
the calibrated spectral-line 
visibilities. Any remaining continuum emission was removed by fitting a first-order polynomial to 
line-free channels in each visibility spectrum, and subtracting this out. The residual visibilities 
were then shifted to the heliocentric reference frame, using the AIPS task CVEL, and then imaged to 
obtain the final spectral cube. The spectra were obtained via a cut through the ``dirty'' cube 
at the location of the target source. 

The spectra of 5 sources from our GPS sample, TXS~1540-077 at $z = 0.172$, TXS~0320+053 at $z = 0.179$, 
TXS~1151-348 at $z = 0.258$, TXS~0019-000 at $z = 0.305$, and TXS~0240-217 at $z = 0.314$, show a clear 
ripple across the observing band. Attempts to excise these ripples by careful data editing were 
unsuccessful. Wide-band RFI features, spanning $\approx 50-80$ channels, were also visible in 
$\approx 70\%$ of the data in these sources, making it impossible to obtain clean spectra. 
We will hence exclude these 5 GPS sources from our sample, and from the later discussion.

\subsection{Results}
\label{results}

The seven GPS sources with RFI-free spectra were found to be compact in the GMRT 
continuum images (angular resolution $\approx 7-10''$) at the respective redshifted 
\hii\ line frequencies. The task JMFIT was hence used to fit a single-Gaussian model 
to a small region around each source, to estimate its flux density. Table~\ref{tab:gps_12} 
lists the measured flux densities of the seven sources. The listed flux densities at the 
redshifted \hii\ line frequencies of the remaining 5 sources, whose data were corrupted 
by RFI, were obtained by interpolating between the 1.4 GHz (from the FIRST or NVSS surveys; 
\citealp[][]{becker1995}; \citealp[][]{condon1998}) and 325/365 MHz (from the WENSS or Texas 
surveys; \citealp[][]{douglas1996}; \citealp[][]{rengelink1997}) flux densities from the literature. 

\begin{figure*}
\begin{tabular}{ccc}
\includegraphics[scale=0.3]{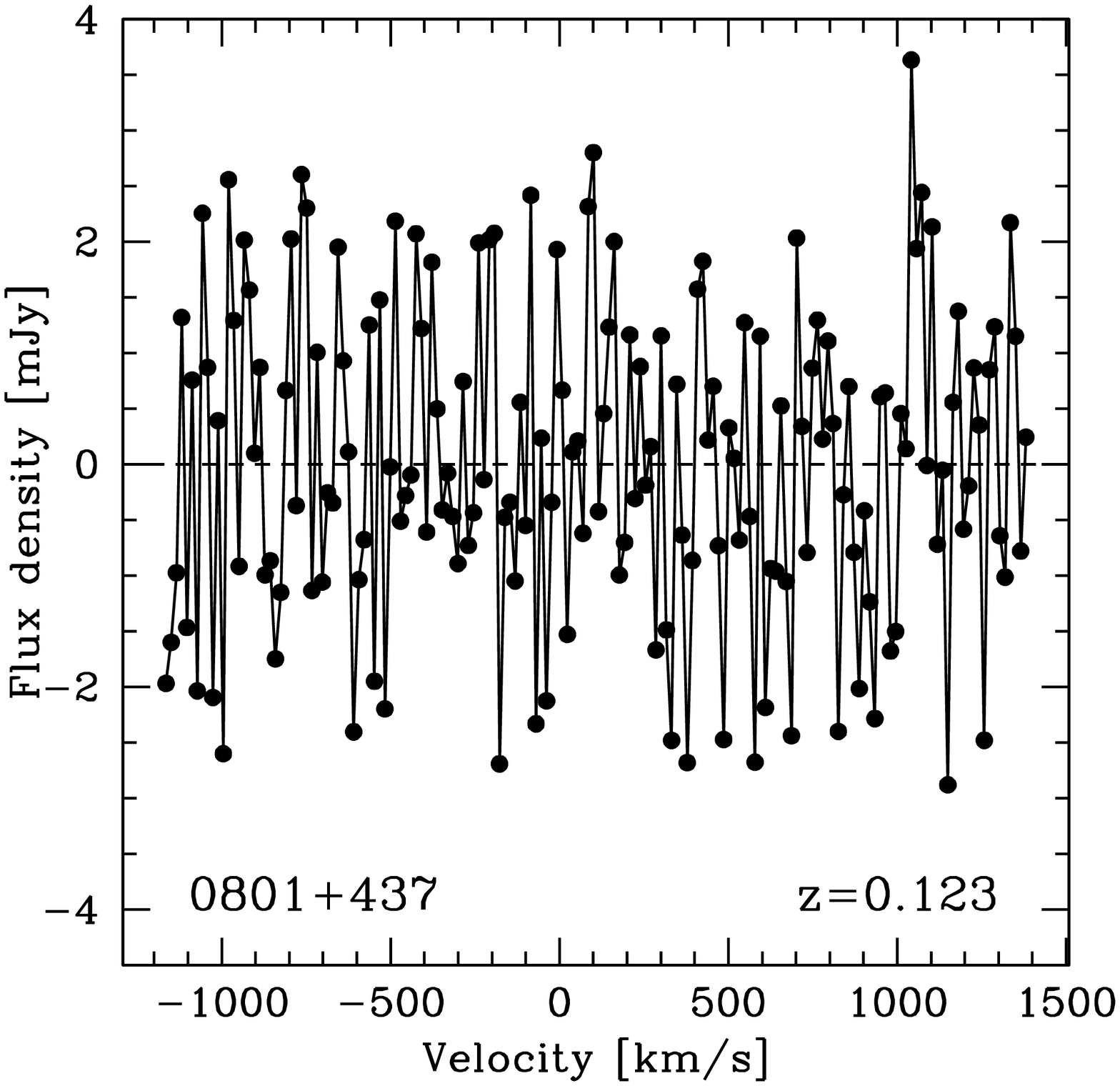} &
\includegraphics[scale=0.3]{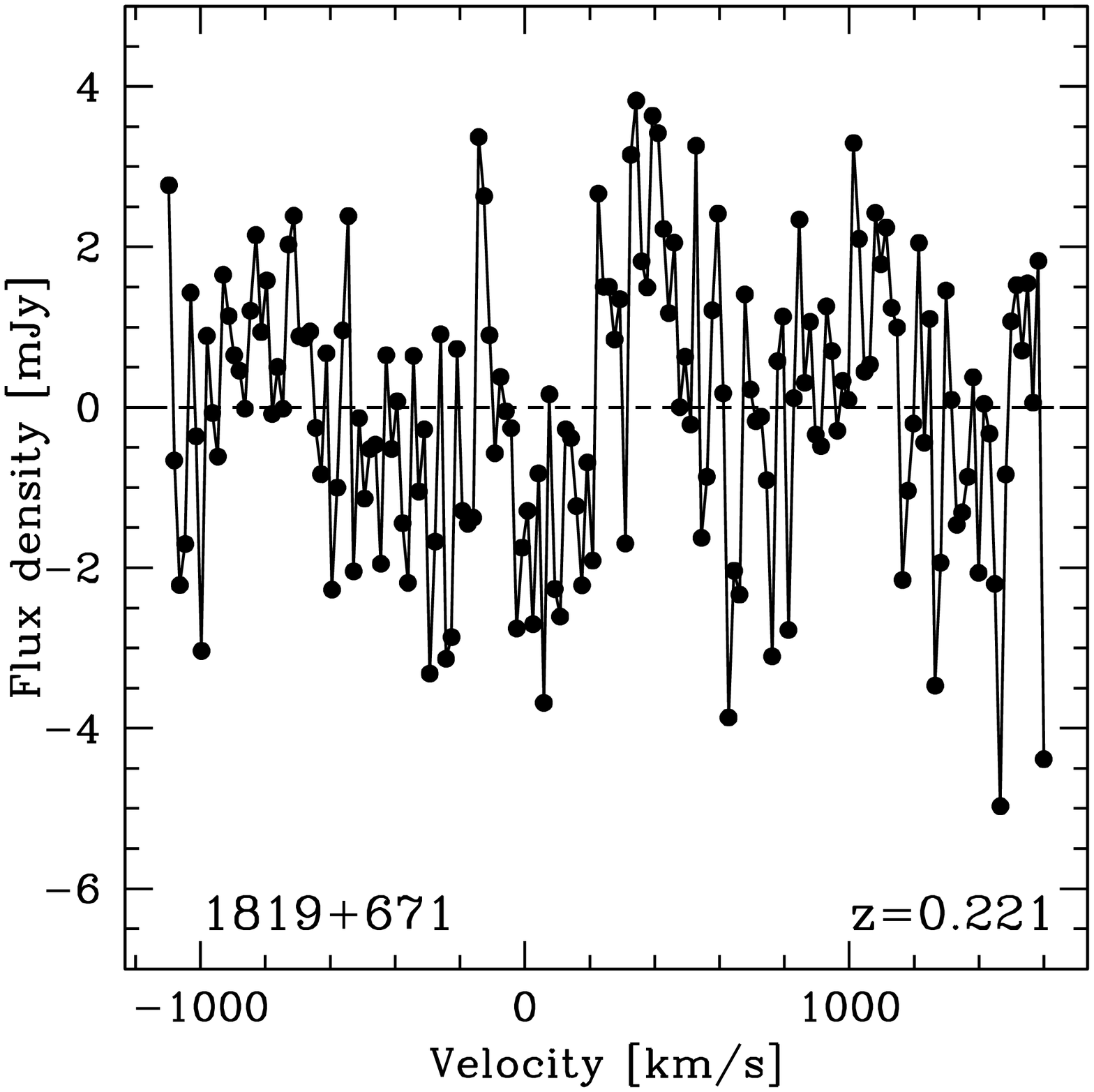} & 
\includegraphics[scale=0.3]{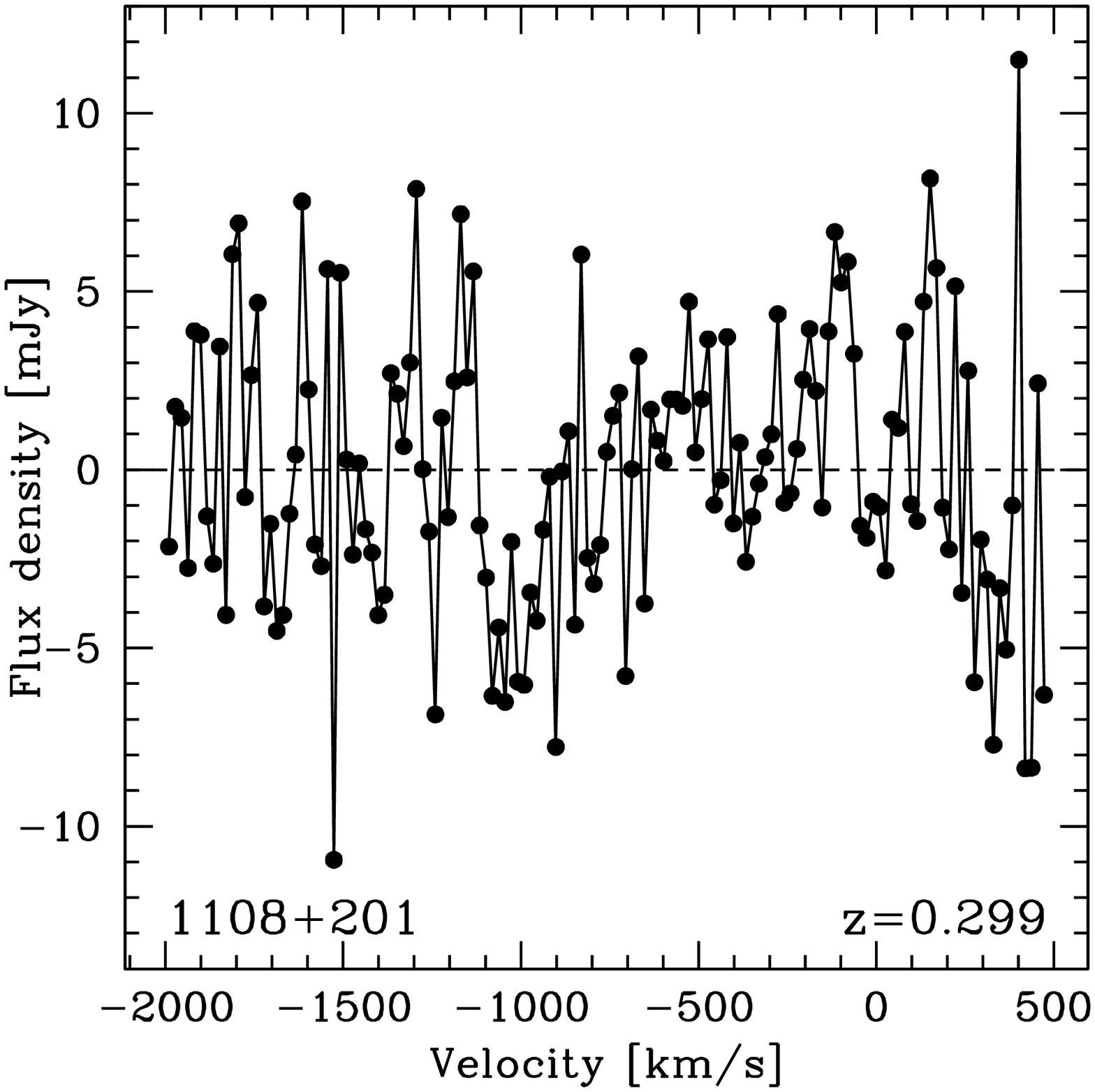} \\

\includegraphics[scale=0.3]{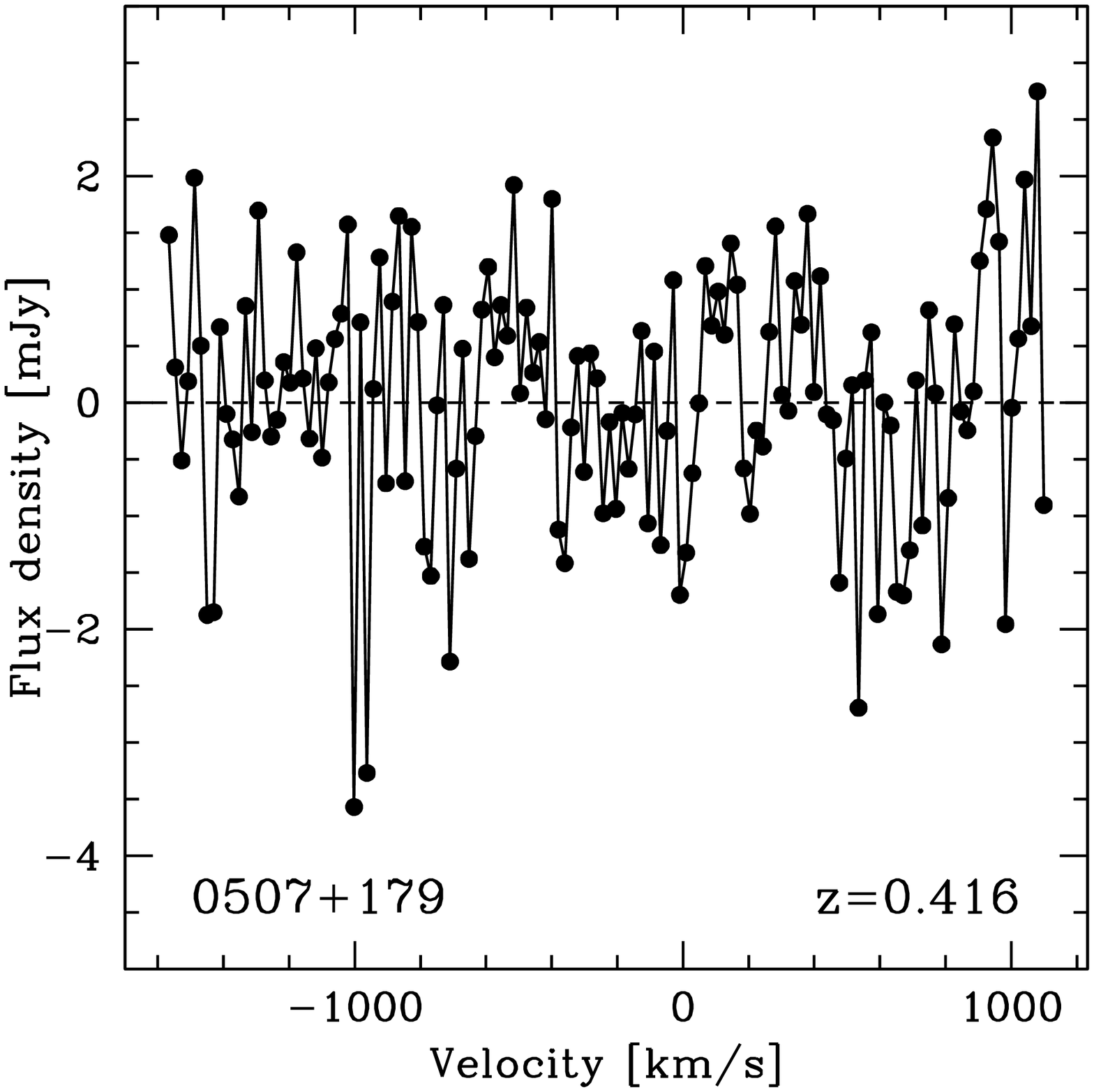} &
\includegraphics[scale=0.3]{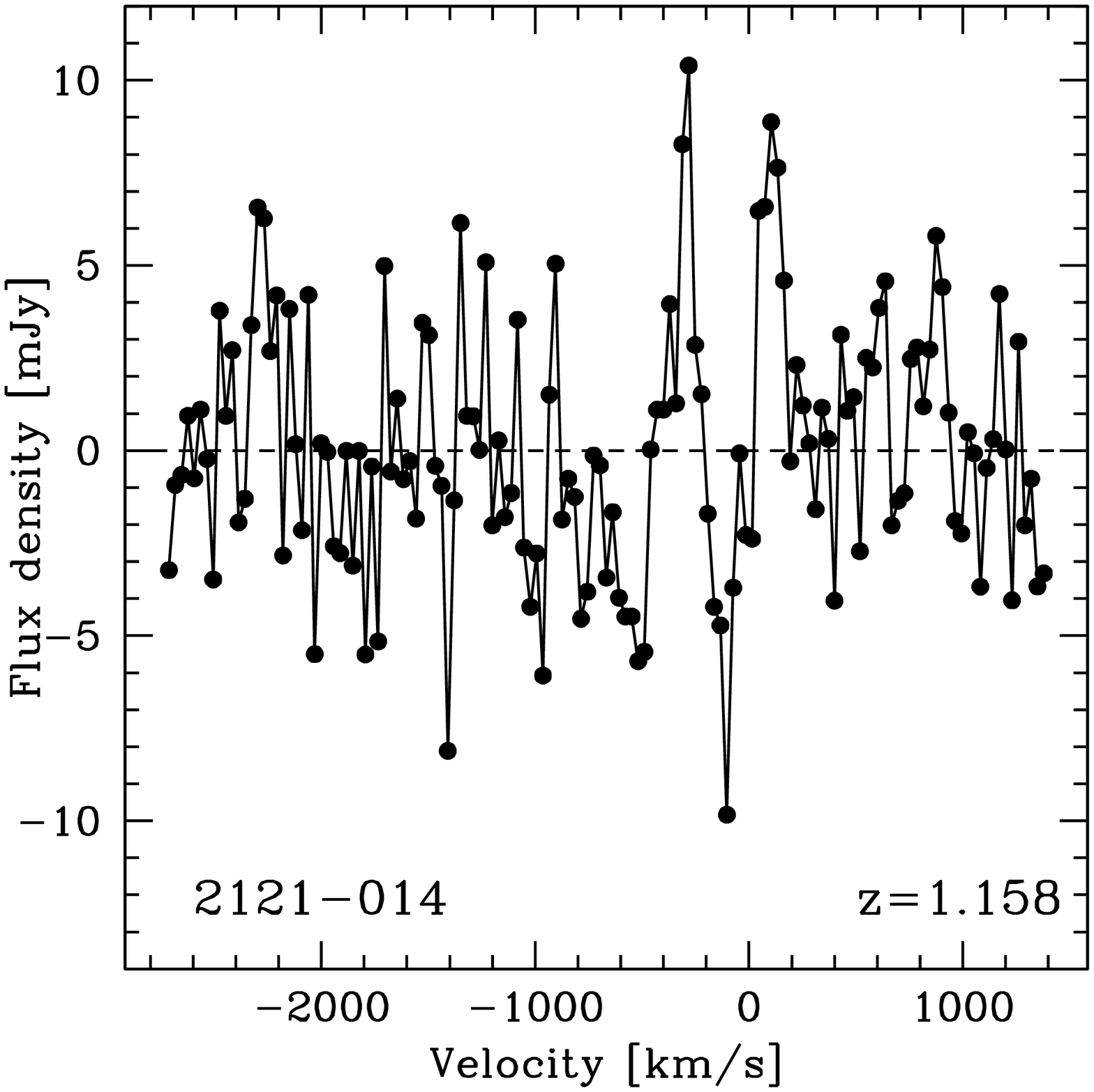} &
                                                      \\
\end{tabular}
\caption[The GMRT \hii\ absorption spectra for the 5 GPS sources]{The GMRT \hii\ absorption spectra of the 
5 GPS sources with non-detections of \hii\ absorption. }
\label{fig:nondetect}
\end{figure*}

We obtained two new detections and 5 non-detections of associated \hii\ absorption from our 7 GPS 
targets. The \hii\ absorption spectra of the detections and non-detections are shown in 
Figures~\ref{fig:detect} and \ref{fig:nondetect}, respectively, in order of increasing redshift. 
All spectra show flux density (in mJy, after subtracting out the source flux density) plotted 
against velocity (in km~s$^{-1}$), relative to the source redshift. The spectra of the non-detections 
in Figure~\ref{fig:nondetect} have been Hanning-smoothed and re-sampled to the velocity resolutions listed
in Table~\ref{tab:gps_12}. For the \hii\ detection towards TXS~1200+045, the spectrum shown in 
Figure~\ref{fig:detect} is from the second observing run in August~2016, which has a higher sensitivity.
For the second \hii\ detection, towards TXS~1245$-$197, the spectrum shown in the figure was obtained 
by combining the spectra from August~2016 and June~2017, with appropriate weights.

Table~\ref{tab:gps_12} summarizes the results from our GMRT observations. The columns of this table are 
(1)~the AGN name, (2)~the AGN redshift, $z$, (3)~the redshifted \hii\ line frequency, $\nu_{21\,cm}$, 
in MHz, (4)~the velocity coverage around the redshifted \hii\ line frequency, in \kms, 
(5)~the velocity resolution, in km~s$^{-1}$, after Hanning-smothing and re-sampling, 
(6)~the AGN flux density, $S_{\nu}$, in mJy, measured at the observing frequency listed in column~(3), 
(7)~the root-mean-square (RMS) noise $\Delta S$ on the final \hii\ spectrum at the velocity resolution of 
column~(5), (8)~the integrated \hii\ optical depth $\int \tau dV$ in km~s$^{-1}$, or, for non-detections, 
the $3\sigma$ upper limit on $\int \tau dV$, (9)~the \hi\ column density $N_{HI}$ in cm$^{-2}$, or, for 
non-detections, the $3\sigma$ upper limit on $N_{HI}$, assuming a gas spin temperature of 1000~K. 
We emphasize that, for non-detections, the upper limits on the integrated \hii\ optical depth and the 
\hi\ column density assume that the line profile has a Gaussian shape with a full-width-at-half-maximum 
of 100~km~s$^{-1}$, with the $3\sigma$ optical depth limit computed at the same velocity resolution.
For detections, the RMS noise values on the final spectra were obtained over line-free channels.
The assumed spin temperature of 1000~K was used for consistency with the literature \citep[e.g.][]{morganti05};
however, we note, in passing, that high spin temperatures are expected for neutral gas located 
close to a bright radio source \citep[e.g.][]{maloney96}.

The two new detections of associated \hii\ absorption are towards TXS~1200+045, at $z = 1.226$, and 
TXS~1245$-$197 at $z = 1.275$. TXS~1200+045 is compact in the GMRT 638~MHz image, with a peak flux density 
of $1675.2 \pm 0.4$~mJy (from JMFIT). The GMRT \hii\ absorption spectrum towards the source from 2016~August 
is displayed in the left panel of Figure~\ref{fig:detect}. The spectrum has two distinct features, narrow deep 
absorption at $\approx -2300$~km~s$^{-1}$ relative to the AGN redshift, and wide, weak absorption extending over 
$\approx 500$~\kms\ over $\approx -2200$~km~s$^{-1}$ to $\approx -1700$~km~s$^{-1}$. Both features were clearly 
detected in the spectra taken at both observing epochs, and in both polarizations.  Both the narrow and the 
wide features seen in the spectrum are hence likely to be real. The integrated \hii\ optical depth of the 
absorption profile is $2.52 \pm 0.12$~km~s$^{-1}$, which implies a high \hi\ column density of 
$\text{N}_\text{HI} = (4.59 \pm 0.22) \times (\text{T}_{s}/1000~\text{K}) 
\times 10^{21} \text{cm}^{-2}$, for an assumed spin temperature of 1000~K and a covering factor of unity. 

TXS~1245$-$197 is also compact in the GMRT 624~MHz image, with a peak flux density of $8302.2 \pm 0.6$~mJy, 
(from JMFIT). The GMRT \hii\ absorption spectrum is displayed in the right panel of Figure~\ref{fig:detect};
this was obtained by averaging the spectra from August 2016 and June 2017. We note that the wide 
absorption feature was detected in both observing runs, with the same depth, and conclude that it is
very likely to be real. The integrated \hii\ optical depth is $4.542 \pm 0.043$~km~s$^{-1}$; this 
implies a high \hi\ column density, $\text{N}_\text{HI} = (8.280 \pm 0.078) \times (\text{T}_{s}/1000~\text{K}) 
\times 10^{21} \text{cm}^{-2}$, for an assumed spin temperature of 1000~K and a covering factor of unity.

\begin{figure*}
\includegraphics[scale=0.28]{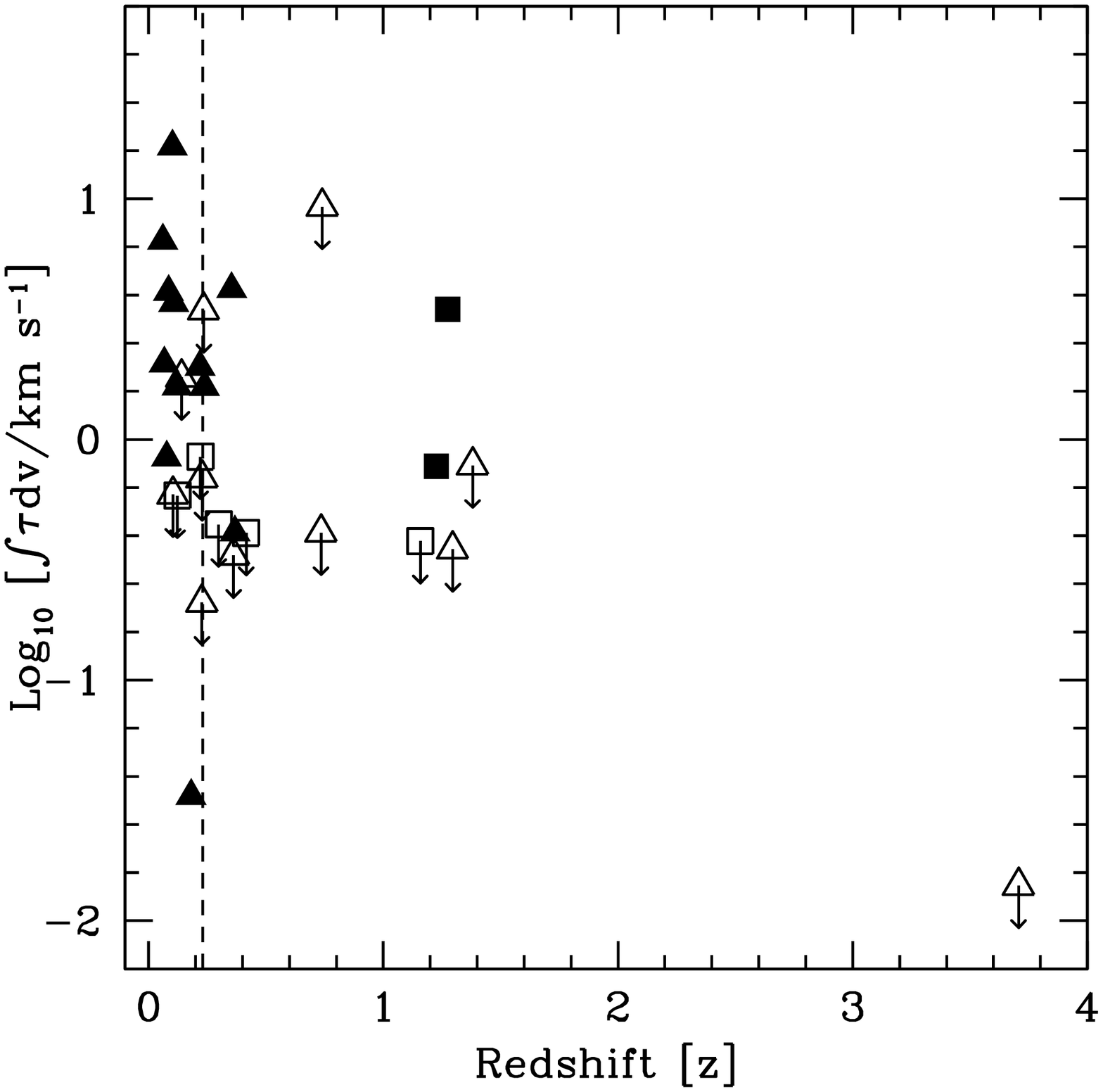}
\includegraphics[scale=0.28]{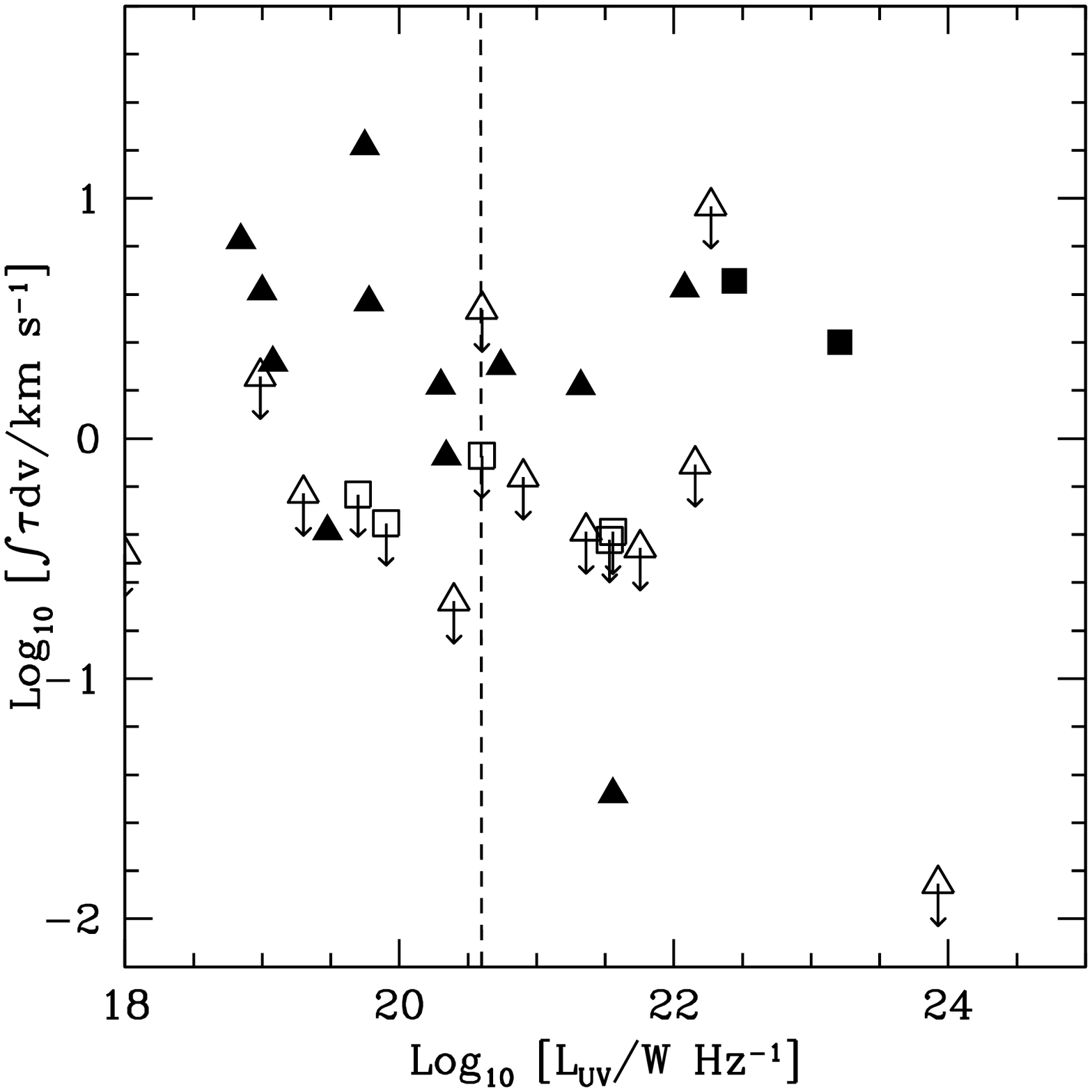}
\includegraphics[scale=0.28]{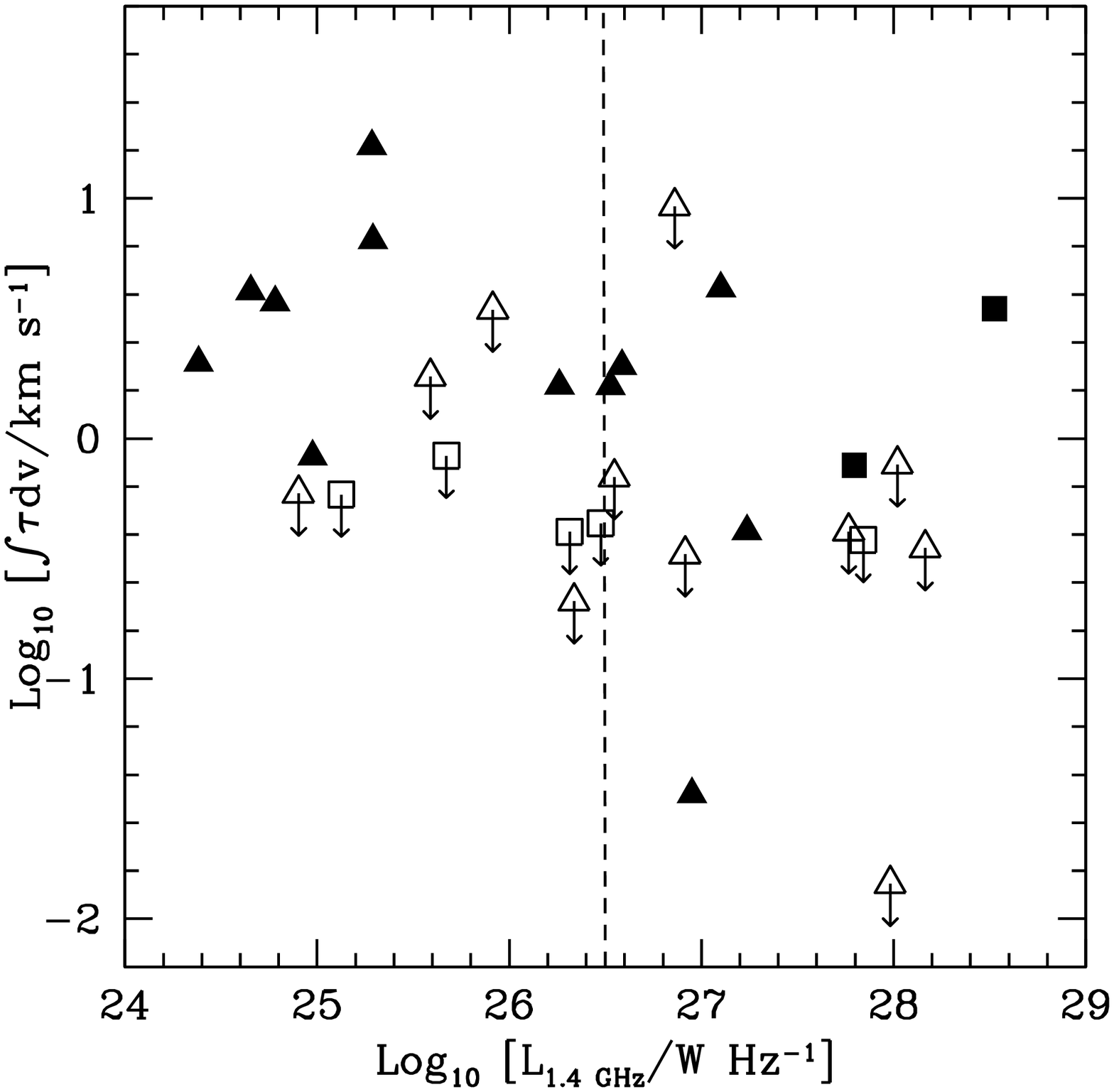}
\caption[]{The integrated \hii\ optical depths of the sample of 30 GPS sources, plotted as a 
function of [A]~redshift, [B]~rest-frame $1216 \AA$ UV luminosity, and [C]~rest-frame 1.4~GHz
radio luminosity. The 7 sources observed by us using the GMRT are shown as squares, while 
the 23 literature sources are shown by triangles. Filled symbols indicate detections of 
\hii\ absorption, while open symbols indicate upper limits on the \hii\ optical depth. 
The dashed vertical line in each panel indicates the median value of the abscissa, 
$z_{med} = 0.23$, ${\rm L}_{UV,med} = 10^{20.6}~W~Hz^{-1}$, and ${\rm L}_{1.4~GHz,med} = 10^{26.5}~W Hz^{-1}$.
}\label{fig:gps}
\end{figure*}

\begin{table*}
\begin{center}
\footnotesize
\caption{The 30 GPS sources with \hii\ absorption searches, including 7 from this study and 23 from the literature, listed in order of 
increasing redshift. The columns of the table are (1)~the AGN name, (2)~the AGN redshift, (3)~the integrated \hi\ 
optical depth, or, for non-detections, the $3\sigma$ upper limit to this quantity, assuming a Gaussian profile with a line 
FWHM of 100~km~s$^{-1}$, (4)~the logarithm of the AGN luminosity $L_{\rm UV}$ (in $W~Hz^{-1}$) at a rest-frame wavelength of 
1216~\AA, i.e. $L'_{\rm UV} = {\rm log} [ L_{\rm UV}/({ W\: Hz}^{-1}) ]$, (5)~the logarithm of the AGN luminosity 
$L_{\rm 1.4\:GHz}$ (in $W~Hz^{-1}$) at a rest-frame frequency of 1.4~GHz, i.e. $L'_{\rm 1.4\: GHz} = 
{\rm log}[ L_{\rm 1.4\: GHz}/({ W\: Hz}^{-1}) ]$, (6, 7)~the two UV or optical wavebands and the measured magnitudes 
therein, that were used to infer the AGN rest-frame 1216~\AA~UV luminosity, (8)~Ref.~\HI, the references for searches 
for associated \hi\ absorption in the literature, and (9)~Ref. UV, the references for the optical/UV fluxes that were 
used to infer the rest-frame 1216~\AA~luminosity. See the main text for details on the computation of the AGN UV luminosity.}
\label{tab:gps}
\begin{tabular}{cccccccccc}
\hline
Source &  $ z $ & $\int \tau dV$~$^a$ & ${\rm L}'_{UV}$~$^b$ & ${\rm L}'_{1.4~GHz}$~$^c$  & Band 1~$^d$   &Band 2~$^d$        & Ref.~$^e$  & Ref.~$^f$   \\
       &        &    km~s$^{-1}$           &                     &                                     &  mag     & mag       & \HI             &    UV           \\

\hline
              TXS 0116+319   & 0.060 &  $5.92 \pm 0.13$   & 18.85 & 25.29 & NUV=20.5 & B=16.7   & 5  & 1,2  \\ 
              B3 1315+415    & 0.066 &  $2.10 \pm 0.33$   & 19.08 & 24.38 & NUV=21.7 & u=20.1   & 4  & 1,3   \\ 
              TXS 1404+286   & 0.077 &  $0.84 \pm 0.16$   & 20.34 & 24.98 & FUV=20.2 & NUV=18.4 & 7  & 1,1   \\ 
	      TXS 0902+468   & 0.085 &  $4.10 \pm 0.51$   & 19.00 & 24.65 & u=18.5   & g=16.6   & 4  & 4,4   \\ 
              TXS 1946+708   & 0.101 &  $15.8 \pm 4.6$    & 19.75 & 25.29 & B=18.3   & R=16.4   & 10 & 2,2     \\ 
              TXS 0729+562   & 0.104 &  $< 0.59$          & 19.30 & 24.90 & B=16.3   & R=13.4   & 4  & 2,2    \\ 
              GB6 J1247+6723 & 0.107 &  $3.70 \pm 0.27$   & 19.78 & 24.78 & FUV=22.9 & NUV=22.6 & 2  & 1,1     \\ 
              TXS 1345+125   & 0.122 &  $1.66 \pm 0.09$   & 20.30 & 26.26 & NUV=19.7 & B=17.0   & 7  & 1,5    \\ 
              B3 0801+437    & 0.123 &  $< 0.58$ 	  & 19.70 & 25.13 & NUV=22.7 & B=17.8   & 11 & 1,2   \\ 
              TXS 1601-222   & 0.141 &  $< 1.8$  	  & 18.99 & 25.59 & B=20.8   & R=18.8   & 7  & 2,2     \\ 
              PKS 1934-63    & 0.181 &  $0.03 \pm 0.01$   & 21.56 & 26.95 & NUV=21.8 & B=17.2   & 6  & 1,2    \\ 
              PKS 0428+20    & 0.219 &  $2.0$   	  & 20.74 & 26.59 & B=20.4   & R=18.6   & 1  & 2,2    \\ 
              TXS 1819+671   & 0.221 &  $< 0.85$	  & 20.60 & 25.67 & NUV=22.1 & B=18.9   & 11 & 1,2   \\ 
              TXS 0941-080   & 0.228 &  $< 0.69$	  & 20.90 & 26.55 & FUV=21.8 & NUV=21.6 & 1  & 1,1    \\ 
              TXS 2021+614   & 0.230 &  $< 0.21$	  & 20.40 & 26.34 & B=19.3   & R=17.5   & 1  & 2,2     \\ 
              TXS 0554-026   & 0.235 &  $< 3.4 $	  & 20.60 & 25.92 & B=18.3   & R=16.5   & 1  & 2,2      \\ 
              TXS 2352+495   & 0.240 &  $1.7$   	  & 21.32 & 26.53 & B=18.3   & V=17.7   & 1  & 8,8     \\ 
              TXS 1108+201   & 0.299 &  $< 0.44$ 	  & 19.90 & 26.48 & u=21.8   & g=20.9   & 11 & 4,4  \\ 
              TXS 2050+364   & 0.354 &  $4.2$   	  & 22.08 & 27.10 & B=18.0   & R=17.2   & 1  & 2,2    \\ 
              TXS 1117+146   & 0.362 &  $< 0.33$	  & 18.01 & 26.92 & u=23.9   & g=21.8   & 1  & 3,3     \\ 
              TXS 1323+321   & 0.368 &  $0.41$ 		  & 18.50 & 27.24 & u=22.9   & g=20.5   & 1  & 3,3    \\ 
              TXS 0507+179   & 0.416 &  $< 0.41$  	  & 21.56 & 26.31 & B=20.2   & R=19.0   & 11 & 2,2   \\ 
              8C 2342+821    & 0.735 &  $< 0.41$          & 21.36 & 27.77 & B=21.4   & R=20.0   & 1  & 2,2     \\ 
              TXS 2149+056   & 0.740 &  $< 9.3 $          & 22.27 & 26.86 & NUV=22.4 & R=20.2   & 9  & 1,7     \\ 
              TXS 2121-014   & 1.158 &  $< 0.38$          & 21.53 & 27.84 & NUV=23.9 & R=23.3   & 11 & 1,5   \\ 
              TXS 1200+045   & 1.226 &  $2.52 \pm 0.12$   & 23.21 & 27.80 & NUV=20.4 & u=18.9   & 11 & 1,3   \\ 
              TXS 1245$-$197   & 1.275 &  $4.542 \pm 0.043$ & 22.44 & 28.52 & NUV=22.1 & B=21.8   & 11 & 1,6   \\ 
              TXS 1518+046   & 1.296 &  $< 0.35 $         & 21.76 & 28.17 & u=23.2   & g=23.1   & 7  & 6,6     \\ 
              TXS 2055+055   & 1.381 &  $< 0.78 $         & 22.16 & 28.02 & NUV=23.2 & R=23.4   & 7  & 1,7    \\ 
              TXS 1351-018   & 3.707 &  $< 0.014$         & 23.93 & 27.98 & g=20.2   & r=19.3   & 8  & 3,3    \\ 
\hline
\end{tabular}
\end{center}
\vskip 0.05in
Notes to the table: \\
$^a$~The integrated \hii\ optical depth or, for \hii\ non-detections, the 3$\sigma$ limit to the \hii\ optical depth,
in km~s$^{-1}$, assuming a line FWHM of 100~km~s$^{-1}$.\\
$^b$~The inferred 1216~\AA\ UV luminosity, obtained by extrapolating from measurements in two nearby optical or UV bands \\
(following the procedure of \citet[][]{curran2008}).\\
$^c$~The rest-frame 1.4~GHz radio luminosities. \\ 
$^d$~The measurements at the two UV/optical wavebands that were used to infer the 1216~\AA~UV luminosities.\\

$^e$~References for the associated \hii\ absorption searches: (1)\citealp[][]{vermeulen2003}; (2)\citealp[][]{saikia2007}; (3)\citealp[][]{pihlstrom2003}; 
(4)\citealp[][]{chandola2011}; (5)\citealp[][]{vangorkom89}; (6)\citealp[][]{veron2000}; (7)\citealp[][]{gupta2006}; (8)\citealp[][]{curran2008}; 
(9)\citealp[][]{carilli1998}; (10) \citealp[][]{peck1999}. (11)~This work. \\

$^{f}$~References for the ultraviolet, optical and infrared band measurements, that were used to obtain the inferred 1216~\AA\ luminosities 
(following the procedure of \citet[][]{curran2010}): (1) \citealp[][]{bianchi2014}, (2) \citealp[][]{monet2003}, (3) \citealp[][]{abazajian2009},  
(4) \citealp[][]{abazajian2005}, (5) \citealp[][]{surace2000}, (6) \citealp[][]{adelman2008}, (7) \citealp[][]{devries2007},  (8) \citealp[][]{zacharias2004}.\\
$^f$~The first and second entries correspond to the 1st and 2nd UV/optical bands, respectively.  \\
Note: The typical uncertainties on measurements in the different UV, optical and near-IR wavebands are 
(1) 0.1 mag (NUV), (2) 0.2 mag (FUV), (3) 0.05 mag (u), (4) 0.3 mag (B), (5) 0.2 mag (V), (6) 0.01 (r), (7) 0.3 mag (R).
\end{table*}

\section{Discussion}

\subsection{TXS~1200+045, $z = 1.226$ }

While TXS~1200+045 is unresolved in our GMRT 638~MHz continuum image, the 1.6~GHz VLBI
image of this source shows a clear three-component structure 
\citep[][]{liu2007}. The three VLBI components have very different flux densities at 1.6~GHz 
\citep[850~mJy, 81~mJy and 35~mJy in components~A, B and C, respectively;][]{liu2007}, with 
component~A containing $\approx 90$\% of the total VLBI flux density. It is not clear whether the three 
VLBI emission components arise from a core and two lobes or from a one-sided core-jet structure. We 
note that component-A, the brightest of the three VLBI components, lies at one end of the 
structure; this may suggest that the source has a core-jet structure.

The GMRT \hii\ absorption profile towards TXS~1200+045 in the left panel of Figure~\ref{fig:detect} 
has a peak line depth of $\approx 20$~mJy. The absorption appears likely to arise against component-A, 
the strongest of the three VLBI components, as the low flux density of the other two components 
\citep[$< 100$~mJy at 1.6~GHz;][]{liu2007} would imply a high \hii\ opacity to produce the observed 
\hii\ absorption.

The deepest \hii\ absorption feature in the GMRT spectrum is significantly blueshifted from 
the AGN redshift of $z = 1.22597 \pm 0.00084$ \citep[e.g.][]{hewett2010}, by $\approx 2300$~\kms. 
We note that the new Data Release-13 of the Sloan Digital Sky Survey (SDSS-DR13) lists the AGN redshift 
as $z=1.22429 \pm 0.00043$, slightly lower than that of \citet{hewett2010}, although consistent within
$2\sigma$ errors. Using the SDSS-DR13 redshift would yield a difference of $\approx 2100$~km~s$^{-1}$ 
between the peak of the \hii\ absorption and the AGN systemic velocity, and would thus not significantly
alter our results. This suggests that the absorption arises in outflowing gas, either driven by a galactic 
wind or mechanically pushed by the radio jet to very high velocities. The secondary feature in the 
spectrum, the wide, weak absorption spanning $\approx 500$~\kms, suggests the presence of disturbed gas, 
possibly due to a jet-cloud interaction. 

In passing, we note that the large velocity offset between the \hii\ absorption and the AGN redshift
raises the possibility that the absorption might arise in an intervening galaxy, rather than 
in gas associated with the AGN. We consider this unlikely because the velocity spread between nulls 
of the \hii\ absorption is $\approx 600$~km~s$^{-1}$, far larger than has ever been seen in an 
intervening galaxy \citep[the \hii\ velocity spreads in intervening high-$z$ absorbers are typically 
$\approx 30-150$~km~s$^{-1}$; e.g.][]{wolfe82,wolfe85,kanekar06,kanekar07,kanekar09b,gupta12,kanekar14c}.
The observed velocity spread in TXS~1200+045 is much more typical of velocity spreads seen in 
associated \hii\ absorbers \citep[e.g.][]{morganti05}.

\subsection{TXS 1245$-$197, $z = 1.275$ }

Our sample of GPS sources was selected based on the criterion that the sources have inverted spectra, 
with the turnover frequency lying between 300~MHz and 5~GHz \citep[e.g.][]{labiano2007}. The inverted 
spectrum in a GPS source is believed to arise due to synchrotron self-absorption in a compact radio 
emission region, probably in the early stages of AGN evolution. TXS~1245$-$197 has an observed turnover 
frequency of $\approx 400$~MHz, while the turnover frequency in the rest frame of the source is 
$\approx 900$~MHz; its spectral characteristics are those of a normal GPS source.

Curiously, the VLBI 2.3~GHz and 8.6~GHz images of TXS~1245$-$197 \citep[][]{sokolovsky2011} show two 
prominent parsec-scale radio lobes, along with a faint extended structure. \citet[][]{sokolovsky2011} found 
the two lobes to have comparable flux densities at both 2.3 and 8.6~GHz, and to have steep spectra between 
the two frequencies, with spectral indices of $-0.72$ and $-0.59$. The authors hence classified the source 
as a Compact Symmetric Object (CSO), as the steep spectra make it unlikely that either source component arises 
from a compact core. CSOs are powerful extragalactic radio sources that show emission on both sides of
an AGN, and have sizes $< 1$~kpc \citep[e.g.][]{wilkinson1994}. Relativistic beaming is believed to be 
small or non-existent in these objects owing to their orientation close to the plane of the sky. Hence, 
the two brightest VLBI source components in TXS 1245$-$197 are likely to correspond to parsec-scale lobes at 
the ends of VLBI-scale jets, with the core remaining undetected at radio frequencies. 

If the steep spectra of the two compact components of the 2.3~GHz and 8.6~GHz VLBI images 
\citep[][]{sokolovsky2011} extend to low frequencies, these components would dominate the 624~MHz 
flux density of TXS~1245$-$197. The radio core is faint at 2.3~GHz and is likely to have an inverted 
or flat spectrum; it is hence unlikely to contribute significantly to the 624~MHz emission detected 
with the GMRT. As such, the detected \hii\ absorption is likely to arise against one or both of the 
radio lobes detected in the VLBI image. The \hii\ absorption profile has a width of $\approx 325$~\kms 
between 20\% points, but has an wide, weak wing extending $\approx 800$~\kms\ blueward of the AGN 
redshift. A wide wing also extends $\approx 300$~\kms\ redward of the AGN redshift, suggesting that 
the neutral gas may be interacting with the AGN jets. The large velocity spread of the blueshifted 
absorption suggests that the bulk of the absorption may arise in outflowing neutral gas from the AGN, 
possibly detected against both radio lobes, as has been observed at lower redshifts 
\citep[e.g.][]{morganti05}. 

\subsection{Mass outflow rates}

Both our \hii\ absorbers show clear evidence for wide, blueshifted absorption, suggestive of outflowing
neutral gas. Assuming that the outflow is driven by a mass-conserving free wind, the mass outflow rate 
${\dot M}$ for neutral gas is given by \citep{heckman02,morganti05}
\begin{equation}
{\dot M} = 30 . \left[\frac{\Omega}{4\pi}\right] .
\left[\frac{r_\star}{1\; {\rm kpc}}\right] .
\left[\frac{{\rm N}_{\rm HI}}{10^{21}\; {\rm cm}^{-2}}\right] .
\left[\frac{V}{300\;{\rm km/s}}\right] \; {\rm M}_\odot \; {\rm yr}^{-1} \;,
\end{equation}
where $V$ is the outflow velocity with which the gas flows into a solid angle $\Omega$ from a minimum raduus
$r_\star$. Following \citet{morganti05}, the solid angle $\Omega$  has been assumed to be $\pi$ radians, 
while the outflow velocity has been assumed equal to half the Full-Width-at-Zero-Intensity (FWZI),
relative to the AGN systemic velocity (i.e. half the velocity width from the outer edge of the blueshifted 
\hii\ absorption to the AGN systemic velocity); this corresponds to $\approx 1200$~\kms\ for TXS~1200+045 
and $\approx 500$~\kms\ for TXS~1245$-$197. The radius is assumed to be equal to half the linear 
size of the detected low-frequency VLBI emission; this has angular extents of $\approx 75$~mas 
\citep[TXS~1200+045; ][]{liu2007} and $\approx 40$~mas \citep[TXS~1245$-$197;][]{sokolovsky2011}, 
corresponding to linear extents of $\approx 458$~pc (TXS~1200+045) and $\approx 340$~pc for TXS~1245$-$197.
We hence use $r_\star \approx 230$~pc (TXS~1200+045) and $r_\star \approx 170$~pc (TXS~1245$-$197).
Note that the derived mass outflow rate also depends critically on the assumed spin 
temperature; our assumed spin temperature of 1000~K allows a direct comparison with the low-$z$ 
results of \citet{morganti05}. This yields mass outflow rates of 
${\dot M} \approx 32 \: {\rm M}_\odot$~yr$^{-1}$ (TXS~1200+045) and ${\dot M} \approx 18 
\: {\rm M}_\odot$~yr$^{-1}$ (TXS~1245$-$197). These lie in  the middle of the range of 
mass outflow rate estimates from similar studies at low redshifts, ${\dot M} \approx 1-50 \: 
{\rm M}_\odot$~yr$^{-1}$ \citep{morganti05}, and comparable to the starburst-driven outflow rates in 
massive ultraluminous infrared galaxies \citep[e.g.][]{heckman02,rupke02}.

\subsection{Dependence on redshift and AGN luminosity }

There are more than 50 detections of associated \hi\ absorption reported in the literature, with the vast majority
of the absorbers detected at low redshifts, $z< 1$ \citep[e.g.][]{vermeulen2003, gupta2006, curran2010, gereb2015}. 
Most searches for associated \hi\ absorption at $z > 1$ have been unsuccessful, with only five detections 
present in the literature. Indeed, we have earlier found evidence that the strength of \hii\ absorption in a 
uniformly-selected sample of compact flat-spectrum sources depends on both redshift and AGN luminosity, with weaker 
\hii\ absorption seen in both the high-redshift and high-luminosity sub-samples \citep{aditya2016}. The weaker 
strength of \hii\ absorption at high redshifts might arise due to redshift evolution of the AGN environment. However,
it is also possible that the observed redshift dependence of the strength of \hii\ absorption arises due to 
an underlying luminosity bias in AGN samples, as a high AGN luminosity in the ultraviolet (UV) or the radio wavebands
can lead to a low \hii\ optical depth \citep[e.g.][]{curran2008}, due to either ionization of the neutral hydrogen (i.e. 
a lower \hi\ column density) or changes in the hyperfine level populations (i.e. a higher spin temperature). Such a 
luminosity bias was indeed present in our sample of flat-spectrum sources, with the high-$z$ AGNs having higher UV and 
radio luminosities \citep{aditya2016}. It is hence not clear whether AGN luminosity or redshift evolution is the primary 
cause of the low observed \hii\ optical depths in high-redshift, high-luminosity active galactic nuclei.

There are 23 GPS sources with searches for associated \hi\ absorption in the literature. We combined our seven targets
with usable \hii\ absorption spectra with these systems to construct a sample of 30 GPS sources, and examined the sample 
for trends with redshift, rest-frame 1216~\AA~UV and rest-frame 1.4~GHz radio luminosities. Table~\ref{tab:gps} summarizes 
the details of the sample. The luminosity at a rest-frame wavelength of 1216~\AA~for each AGN was estimated 
following the prescription of \citet[][]{curran2008}. We first determined the flux density $F_{UV}$ at the wavelength 
$1216 \times (1 + z)$~\AA~for each AGN, by using a power-law spectrum to interpolate between its measured flux densities 
at two optical and/or UV wavebands in the literature. The luminosity at the rest-frame wavelength of $1216$~\AA~was 
then inferred from the expression ${\rm L}_{UV} = 4\pi D^{2}_{AGN} F_{UV} / (1 + z)$, where $D_{AGN}$ is the luminosity 
distance of the AGN.

We note, in passing, that the sample of GPS sources is now formally sufficiently large (30 systems, with 
10 detections of \hii\ absorption) to examine the dependence of the detection rate of \hii\ absorption on the above 
quantities, especially if all \hii\ detections are obtained at high or low values of the quantity under study. The three 
panels of Figure~\ref{fig:gps} plot the integrated \hii\ optical depth of the 30 GPS sources of the sample against 
redshift, rest-frame $1216$~\AA~UV luminosity, and rest-frame 1.4~GHz radio luminosity. It is clear that the bulk of the 
sample is at very low redshifts and low luminosities. Indeed, the median redshift of the GPS sample is $z_{med} = 0.23$, 
while the median UV and radio luminosities are about an order of magnitude lower than the corresponding median
values for the compact flat spectrum sample of \citet{aditya2016}. Using a Peto-Prentice two sample test, and dividing 
the sample at the median value of each quantity, we find no statistically-significant evidence for a dependence of the 
\hii\ absorption strength on redshift, rest-frame $1216$~\AA~UV luminosity or rest-frame 1.4~GHz radio luminosity.
However, we emphasize that our present sample of GPS sources is still relatively small, and, more important, is 
dominated by low-redshift, low-luminosity active galactic nuclei. Studies of the redshift evolution and luminosity 
dependence of the strength of \hii\ absorption will require significantly larger samples of GPS sources at high 
redshifts, $z > 1$. 


\subsection{Conclusions}

We have used the GMRT 610~MHz receivers to detect redshifted \hii\ absorption from neutral gas associated with 
two high-$z$ GPS sources, TXS~1200+045 at $z = 1.226$, and TXS~1245$-$197 at $z = 1.275$; we also report 5 
new non-detections of \hii\ absorption at $z\approx 0.12-1.2$. The two new \hii\ absorbers have line profiles 
consisting of two components, narrow, deep absorption, and a wide, weak wing, with a significant part of the 
detected absorption blueshifted from the AGN systemic velocity, by $\approx 2300$~km~s$^{-1}$ for TXS~1200+045 
and by $\approx 1000$~km~s$^{-1}$ for TXS~1245$-$197.  In both cases, the blueshifted absorption is likely 
to arise from outflowing neutral gas, perhaps being driven out from the AGN environment by a galactic wind 
or by the radio jets. Assuming a simple wind-driven model for the outflows, we obtain relatively high mass 
outflow rates for the two systems, ${\dot M} \approx 18-32 \: {\rm M}_\odot$~yr$^{-1}$, assuming a gas 
spin temperature of $1000$~K.

Including 23 GPS sources that have earlier searches for associated \hii\ absorption available in the literature, 
we have constructed a sample of 30 GPS sources, and examine the trends of \hii\ absorption strength with 
redshift, rest-frame 1216~\AA~UV and rest-frame 1.4~GHz radio luminosity. Unlike in the case of earlier \hii\ 
absorption studies of samples of compact flat-spectrum sources, we do not find statistically-significant 
evidence for the hypothesis that the strength of \hii\ absorption in our sample depends on redshift or 
rest-frame AGN 1216~\AA~UV or 1.4~GHz~radio luminosity. This is likely to be because our GPS sample is both 
relatively small and dominated by low-redshift, low-luminosity active galactic nuclei.

\section*{Acknowledgements}
We thank the staff of the GMRT who have made these observations possible. The GMRT is run by the National Centre 
for Radio Astrophysics of the Tata Institute of Fundamental Research. NK acknowledges support from the 
Department of Science and Technology via a Swarnajayanti Fellowship (DST/SJF/PSA-01/2012-13). We thank 
an anonymous referee for a very helpful report on an earlier draft that significantly improved the 
quality of this paper.


\bibliographystyle{mnras}
\bibliography{ms}

%

\bsp	
\label{lastpage}
\end{document}